\documentclass[11pt,superscriptaddress,aps,prd,preprint]{revtex4}
\usepackage[active]{srcltx}
\usepackage{graphicx}
\usepackage[utf8]{inputenc}
\usepackage{amsmath}
\usepackage{xcolor}
\usepackage{times,txfonts}
\begin{document}

\title{A New Braneworld with Conformal Symmetry Breaking}

\author{G. Alencar}
\affiliation{Universidade Federal do Cear\'{a}, Fortaleza, Cear\'{a}, Brazil}
\email{geova@fisica.ufc.br}

\author{I. C. Jardim}
\affiliation{Departamento de F\'{i}sica, Universidade Regional do Cariri, 57072-270, Juazeiro do Norte, Cear\'{a}, Brazil}
\email{ivan.jardim@urca.br}

\author{R. I. de Oliveira Junior}
\affiliation{Faculdade de Educação, Ciências e Letras do Sertão Central(FECLESC), Rua José de Queiroz Pessoa- Planalto Universitário, Quixadá-CE, 63900-000, Brazil}
\email{raimundo.ivan@uece.br}

\author{M. Gogberashvili}
\affiliation{Ivane Javakhishvili Tbilisi State University, 3 Chavchavadze Avenue, Tbilisi 0179, Georgia}
\email{gogber@gmail.com}

\author{R. N. Costa Filho}
\affiliation{Universidade Federal do Cear\'{a}, Fortaleza, Cear\'{a}, Brazil}
\email{rai@fisica.ufc.br}

\begin{abstract}
We explore the conformal 5D braneworld, where warping emerges through conformal symmetry breaking. Our scenario seamlessly aligns with conventional brane approaches if conformal symmetry remains unbroken. It is shown that a model with a single conformal breaking parameter effectively localizes gravity on the brane, but it falls short in trapping gauge bosons. However, in scenarios with two parameters, gravity is localized, and the model also achieves the localization of zero modes for both gauge and Dirac fields.

\end{abstract}
\maketitle

\section{Introduction}

The study of conformal transformations in physics traces its roots back to the late 19th and early 20th centuries. Since then, numerous pivotal contributions have shaped our understanding, including the establishment of the invariant form of physical equations by Hermann Weyl \cite{Kastrup:2008jn}, the construction of the conformal group in Special Relativity \cite{DiFrancesco:1997nk, Deriglazov:2004yr} (attributed to Weyl and Cartan), and Weyl's gauge symmetry \cite{Ghilencea:2018thl}. Further advancements include their application in Quantum Field Theory \cite{Alhaidari:1987idh}, Conformal Field Theory \cite{Gillioz:2022yze, Qualls:2015qjb}, and String Theory \cite{Chanson:2022wls, Budzik:2020aqg}.

Conformal transformations are pivotal in physics, offering valuable insights and simplifications across various theoretical frameworks. They contribute to our understanding of Critical Phenomena and Phase Transitions \cite{Hongler:2013veg}, the Quantum Hall Effect \cite{Gaite:1994xp, Cappelli:1992kf}, Black Hole Physics \cite{Xu:2023tfl, Atkins:2023axs}, the Renormalization Group \cite{Schafer:1976ss}, and Cosmology \cite{Faraoni:1998qx, Carloni:2010rfq}.

In the realm of high-dimensional physics, particularly in string theory and specific areas of cosmology, conformal symmetries and transformations are also explored. Notably, the Ads/CFT correspondence leverages the conformal symmetry of the boundary theory for its duality \cite{Papadimitriou:2004ap,Coleman:2020jte,Samantray:2013ada}.

Symmetry breaking is as crucial as symmetry itself. The breaking of conformal symmetry assumes particular importance in Phase Transitions and Critical Phenomena. Introducing a relevant scale breaks conformal symmetry, giving rise to distinct phases in physical systems \cite{Maldacena:2016upp, Chankowski:2014fva}. This breaking also plays a pivotal role in the Higgs mechanism within the Standard Model of particle physics. In this context, the breaking of conformal symmetry leads to the masses of elementary particles \cite{Mannheim:2016lnx,Karch:2023wui}.

In cosmology, understanding the breaking of conformal symmetry is crucial for comprehending the universe's evolution. For instance, breaking the scale invariance associated with conformal symmetry during inflationary periods can generate structure in the universe \cite{A.B Arbuzova, Ghilencea:2021jjl}. In the AdS/CFT correspondence, breaking conformal symmetry on the boundary corresponds to a gravitational background (AdS space) with a scale in the bulk. Insight into how conformal symmetry is broken on the boundary provides valuable perspectives on the dual gravitational description in the bulk \cite{Hull:2022vlv, Arefeva:1998cvg, Igor R. Klebanov}.

In the context of high-dimensional physics, intriguing scenarios emerge where our observable universe is associated with a brane embedded in a higher-dimensional space-time exhibiting non-factorizable geometry \cite{ArkaniHamed:1998rs, Gogberashvili:1998vx, Randall:1999vf}. In such configurations, a key requirement for realizing the braneworld concept is the localization of various matter fields on the brane (see the reviews \cite{Langlois:2002bb, Mannheim, Maartens:2010ar} for in-depth discussions). The quest for a universal trapping mechanism that accommodates all fields is preferred. However, most braneworld models encounter difficulties in achieving such a mechanism \cite{Bajc:1999mh, Chang:1999nh, Davoudiasl:1999tf, Pomarol:1999ad, Gogberashvili:2003xa, Jardim:2014xla, Alencar:2014moa}.

Recently, a new braneword scenario has been proposed based on a hidden conformal symmetry of the 5D brane model \cite{Alencar:2017dqb}. In this model, the scalar field (the conformon) or the warp factor can effectively localize gravity on the brane. However, they face challenges in trapping gauge or Dirac fields unless non-minimal couplings with gravity are introduced. This paper explores an extension of the model through a slight breaking of the conformal symmetry in the conformon potential. This modification allows us to discover new solutions for the scalar field and the warp factor, addressing the localization problem for gravity, gauge, and Dirac fields.

The manuscript is organized as follows: Section 2 outlines the setup of our 5D model. Sections 3-5 detail the localizations of gravity, gauge fields, and fermions on the brane.


\section{The Background}

Consider the 5D action featuring a conformally coupled scalar field, denoted as $\chi$ or conformon:
\begin{equation} \label{scalargravity}
S_{G} = \int\sqrt{-g} d^{5}x \left[\xi\chi^{2}R + \frac{1}{2}\partial_{M}\chi\partial^{M}\chi - U(\chi) - \lambda(\chi)\delta(z)\right]~, \qquad (M,N,... = 0,1,2,3,4)
\end{equation}
where $R$ represents the 5D Ricci scalar, $z$ denotes the extra spatial coordinate, and  $\xi$ is a constant. In (\ref{scalargravity}) the potential functions of $\chi$ are taken in the form:
\begin{equation} \label{U,lambda}
U(\chi) = u_{0}\chi^{m}~, \qquad \lambda(\chi) = \mu\chi^{n}~,
\end{equation}
with constants $m$, $n$, and parameters $\mu$ and $u_{0}$. The action (\ref{scalargravity}) retains invariance under conformal rescalings,
\begin{equation} \label{tilde-chi}
\tilde{g}_{MN} = e^{2\rho (x^M)}g_{MN}~, \qquad \tilde{\chi} = e^{-\frac{3}{2}\rho (x^M)}\chi~,
\end{equation}
with specific parameter values 
\begin{equation} \label{m,n}
\xi = \frac {3}{32}~, \qquad m = n = \frac{10}{3}~.
\end{equation}

The 5D gravitational and conformon field equations associated with (\ref{scalargravity}) are given by:
\begin{eqnarray}
2\xi\chi^{2}G_{MN} & = & T_{MN} ~, \label{Ein01}\\
\Box\chi-2\xi R\chi & = & -\frac{\partial U(\chi)}{\partial\chi}-\frac{\partial\lambda(\chi)}{\partial\chi}\,\delta(z)~, \label{eqChi01}
\end{eqnarray}
where:
\begin{equation}
\begin{split}
 T_{MN} &= - \partial_{M}\chi\partial_{N}\chi + g_{MN}\left[ \frac{1}{2}\partial_{P}\chi\partial^{P}\chi-U(\chi)\right] - 2\xi \left[ g_{MN}\square(\chi^{2})-\partial_{M}\partial_{N}(\chi^{2})\right] - \\
 &- e^{2A}\delta_{M}^{\mu}\delta_{N}^{\nu}g_{\mu\nu}\lambda(\chi)\delta(z)~. \qquad \qquad \qquad \qquad \qquad \qquad \qquad (\mu,\nu,... = 0,1,2,3)
\end{split}
\end{equation}

Now, we want to partially break the conformal symmetry consider
\begin{equation}
\chi = \chi(z)~,
\end{equation}
and assume $\rho = \rho(z)$ in (\ref{tilde-chi}). The action (\ref{scalargravity}) retains a residual conformal symmetry, taking the form:
\begin{equation} \label{action}
S_{G} = \int \sqrt{-g} d^{5}x \left[ \xi\chi^{2}(z)R - \frac{1}{2}\chi\left(\chi'' + 4A'\psi'\right) - U(\psi) - \lambda(\chi)\delta(z) \right] ~,
\end{equation}
where primes denote derivatives with respect to $z$, and we utilize the conformally warped metric:
\begin{equation} \label{metric}
ds^{2} = e^{2A(z)}g_{MN}dx^{M}dx^{N}.
\end{equation}


\section{Localization of gravity}

Given that the Ricci scalar in (\ref{scalargravity}) is multiplied by two functions of $z$, achieving the localization of gravity on the brane at $z = 0$ is possible through either the warp factor $e^{2A(z)}$ or the scalar field $\chi (z)$ \cite{Alencar:2017dqb, Alencar:2017vqd}. To elucidate the localization mechanism with the residual conformal symmetry, let's rewrite the equations of motion (\ref{Ein01}) and (\ref{eqChi01}) as follows:
\begin{eqnarray}
\frac{9}{8}\chi^{2}A'^{2} &=& -\frac{1}{2}\chi'^{2}-e^{2A}u_{0}\chi^{m}~, \label{Ein02}\\
\frac{9}{16} \chi^ {2} \left( A'^{2} + A''\right) &=& \frac{1}{16} \left(5\chi'^{2} -3\chi \chi'' \right) - e^{2A}u_{0}\chi^{m}  - e^{2A} \lambda\chi^{n} \delta(z)~, \label{Ein03}\\
\chi'' + 3A'\chi' + \frac{3}{4}\left(3A'^{2} + 2A''\right)\chi &=& -u_{0}me^{2A}\chi^{m-1} - \lambda n\chi^{n-1}\delta(z)~,\label{eqChi02}
\end{eqnarray}
which correspond to the action (\ref{action}). Introducing the new field,
\begin{equation} \label{chi=}
\psi = e^{\frac{3}{2}A} \chi ~,
\end{equation}
the equation for the conformon (\ref{eqChi02}) simplifies to:
\begin{equation}\label{eqpsim}
\psi'' = -u_{0}me^{-\frac{3}{2}\left(m -\frac{10}{3}\right)A }\psi^{m-1} - \lambda n\psi^{n-1}\delta(z) ~.
\end{equation}
For the specific value of $m$ defined in (\ref{m,n}), the warp factor disappears, yielding the solution:
\begin{equation} \label{psi}
\psi = \frac {C}{(k|z|+1)^{3/2}}~,
\end{equation}
where $C$ and $k$ are integration constants related to the free parameters of the model:
\begin{equation}
u_{0} =-\frac {9k^{2}}{8C^{4/3}}~, \qquad \mu = \frac {9k}{10C^{4/3}}~.
\end{equation}
Using (\ref{chi=}) and (\ref{psi}), the equation for the warp factor (\ref{Ein02}) can be expressed as:
\begin{equation}
 A'\left(A' + \frac{k\,\text{sgn}(z)}{k|z| + 1}\right) = 0~,
\end{equation}
where $\text{sgn}(z)$ denotes the sign function. This equation admits two solutions for the warp factor:
\begin{equation} \label{warp}
(i).~ e^{A} = 1 ~, \qquad (ii).~ e^{A} = \frac {1}{k|z|+1}~,
\end{equation}
both of which satisfy the equation (\ref{Ein03}). In the first case (i), the scalar field $\chi$ is accountable for localizing gravity on the brane. In the second case (ii), the conformon field reduces to a constant, and gravity is localized by the warp factor, as in standard brane scenarios \cite{Gogberashvili:1998vx, Randall:1999vf}. However, to localize gauge fields as well, additional couplings need to be introduced \cite{Alencar:2017dqb}.

The localization of gauge fields on the brane is facilitated if the solution of (\ref{eqpsim}) behaves as $\psi \sim e^{A}$. Achieving this requires setting:
\begin{equation} \label{m}
m = \frac{10}{3} + \alpha~,
\end{equation}
where $\alpha$ is the conformal invariance breaking parameter. Plugging (\ref{m}) into (\ref{Ein01}) and (\ref{eqpsim}), along with $n = \frac{10}{3} + \sigma$ (where $\sigma$ is another constant, not crucial for the model), yields the following set of equations:
\begin{equation} \label{system}
\begin{split}
\beta A'' + \beta^{2}A'^{2} &= -\left(\frac{10}{3}+\alpha\right) u_{0}C^{\frac{4}{3} + \alpha}e^{\gamma A} - \left(\frac{10}{3}+\sigma\right)\mu C^{\left(\frac{4}{3} + \sigma\right)}\delta(z) ~, \\
\left[ \frac{9}{8} + \frac{1}{2} \left( \beta - \frac{3}{2} \right)^{2} \right] A'^{2} &= - u_{0} C^ {\frac{4}{3} + \alpha} e^ {\gamma A}~,
\end{split}
\end{equation}
where
\begin{equation} \label{gamma}
\gamma = \frac{4}{3} \beta + \beta \alpha - \frac{3 \alpha}{2}~.
\end{equation}
Solving this system yields the solution:
\begin{equation}\label {b}
A(z) = \ln(\kappa|z|+1)^{-\frac{2}{\gamma}}~,
\end{equation}
and the constraints on the parameters:
\begin{eqnarray}
\left[ \frac{9}{2} + 2 \left( \beta - \frac{3}{2} \right)^{2} \right] \frac{k^{2}}{\gamma^ {2}} &=& - u_{0} C^{\frac{4}{3} + \alpha}~,\label{rela1} \\
-\left( \frac{10}{3} + \alpha \right) u_{0} C^{\frac{4}{3} + \alpha} &=& \left( \frac{2\beta}{\gamma} + \frac{4\beta^{2}}{\gamma^ {2}} \right) k^{2} ~, \label{rela2} \\
\frac{4k\beta}{\gamma} &=& \left(\frac{10}{3}+\sigma \right)\mu C^{\left(\frac{4}{3} + \sigma\right)} ~.\label{rela3}
\end{eqnarray}
From the first two constraints and from the definition (\ref{gamma}), we can obtain the relations:
\begin{equation} \label{beta,gamma}
\beta = \frac{3(10+3\alpha)}{20 + 3\alpha}~, \qquad \gamma = \frac{9\alpha^{2} +24\alpha +80}{2(3\alpha +20)}~.
\end{equation}
It is evident that the parameter $\gamma$ is positive for $\alpha > -20/3$, and $\beta$ is positive for $\alpha > -10/3$ or $\alpha < -20/3$. Assuming a small conformal symmetry breaking, we consider $\beta$ and $\gamma$ as positive constants. When the conformal symmetry breaking parameter in (\ref{m}) vanishes ($\alpha=0$), the solution (\ref{b}) reduces to the standard ones (\ref{warp}), i.e., $\beta = \frac{3}{2}$ and $\gamma = 2$.

The relations (\ref{beta,gamma}) can be employed in equations (\ref{rela2}) and (\ref{rela3}) to establish connections between the parameters $C$ and $k$ with the free parameters of the model, namely, $u_{0}$, $\mu$, and $\alpha$. The explicit relation is not crucial for the outcomes of this study, and its computation is not pursued. However, it's worth noting that for a real and positive value of $k$, it is necessary to fix $\mu > 0$ and $u_{0} < 0$. As mentioned earlier, the parameter $\sigma$ does not impact gravity localization; it merely alters the relationship between $C$ and $k$ in (\ref{rela3}).

We have two additional constraints to impose on the model, both stemming from the requirement to recover 4D gravity. The condition for gravity to be localized on the brane is given by:
\begin{equation}
\int\frac{C^{2}}{(\kappa|z|+1)^{\frac{4\beta}{\gamma}}}dz= \frac{2C^ {2}}{k\left(  \frac{4\beta}{\gamma} - 1 \right) } = G~,
\end{equation}
where $G$ is the Newton constant. This condition necessitates $4\beta >\gamma$. Consequently, the constraints provided by (\ref{beta,gamma}) on the value of the conformal symmetry-breaking parameter are:
\begin{equation}
\frac {4\left( 2 - \sqrt{14}\right)}{3} < \alpha < \frac {4\left( 2 + \sqrt{14}\right)}{3}~.
\end{equation}
The case where $\alpha$ takes the value of zero, corresponding to a conformally symmetric model that ensures the localization of gravity, falls within this range. Thus, our background is consistent and benefits from the existence of a new scalar field,
\begin{equation} \label{chi}
\chi = Ce^{(\beta -3/2)A}~,
\end{equation}
which is solely a function of the warp factor $A(z)$. This field allows us to localize standard model fields on the brane. 


\section{Localization of gauge fields}

The Lagrangian for a gauge field in a D-dimensional conformal background (\ref{metric}) can be expressed as \cite{Alencar:2017dqb}:
\begin{equation}
L = \chi^{2 \frac{D-4}{D -2}} F^{2}~.
\end{equation}
For the specific case considered in this paper, $D=5$, and for the conformon, we use the solution (\ref{chi}):
\begin{equation} \label{chi+}
\chi \sim  e^{3 \theta A/2}~,
\end{equation}
where:
\begin{equation} \label{theta}
\theta = \frac{2}{3} \beta - 1~.
\end{equation}
Subsequently, the 5D action for the gauge field takes the form:
\begin{equation}
S = \int d^{5}x \sqrt{-g} e^{\theta A} g^{MP}g^{NQ} F_{MN}F_{PQ}~,
\end{equation}
where $F_{MN} = \partial_{M}{\cal A}_{N} - \partial_{N} {\cal A}_{M}$. This action leads to the following equation for the gauge field ${\cal A}_{N}$:
\begin{equation} \label{AM}
\begin{split}
\partial_{\nu}[ \sqrt{-g} e^{\theta A} F^{5\nu}] &= 0 ~,\\
\partial_{5} [ \sqrt{-g} e^{\theta A} F^{\mu 5} ] + \partial_{\nu}[ \sqrt{-g} e^{\theta A} F^ {\mu\nu}] &= 0 ~.
\end{split}
\end{equation}
Imposing the gauge,
\begin{equation}
{\cal A}_{5} = \partial_{\mu} {\cal A}^{\mu} = 0~,
\end{equation}
and employing the separation of variables,
\begin{equation}
{\cal A}^{\mu}(x,z) = \Tilde{\cal A}^{\mu} (x) f(z)~,
\end{equation}
the system (\ref{AM}) gives the following set of equations:
\begin{equation}
\begin{split}
\Box \Tilde{\cal A}^{\mu} + m^{2}\Tilde{\cal A}^{\mu} &= 0~, \\
f'' (z) + (\theta + 1) A'f' (z) &= m^{2} f(z) ~,
\end{split}
\end{equation}
where $m$ is the separation constant, which can be interpreted as the mass of the 4D gauge field. For $m=0$, the equation for $f(z)$ has the zero-mode solution, $f(z) = C_{0}$, and the effective action of the model takes the form:
\begin{equation}\label{Seff}
S_{\rm eff}= - \frac{C_0^{2}}{4} \int^{\infty}_{-\infty} dz \, e^{A(\theta + 1)} \int d^{4}x \Tilde{F} ^ {\mu\nu} \Tilde{F}_{\mu\nu}~.
\end{equation}
In this expression, the first integral can be written as:
\begin{equation}
I = 2 \int_{0}^{\infty} \frac{1}{(kz + 1)^{2(\theta + 1)/\gamma}} dz~.
\end{equation}
To localize the zero mode, this integral must be convergent, leading us to the condition:
\begin{equation}\label{cond1}
2(\theta + 1) > \gamma ~.
\end{equation}

Using the definitions (\ref{beta,gamma}) and (\ref{theta}), one can show that the relation (\ref{cond1}) can be satisfied only if $\alpha^{2} < 0$, which cannot be achieved for any $\alpha \in \mathbb{R}$. This means that for the conformon model with the single conformal symmetry breaking parameter $\alpha$, the gauge field cannot be localized. But we can introduce the second conformal symmetry breaking parameter $\delta$, and modify the conformon solution (\ref{chi+}) to $\chi^{\frac{2}{3} + \delta}$. This modification is equivalent to the change $\theta \to \theta + \delta(\beta -3/2)$. Then, the localization condition (\ref{cond1}) transforms into:
\begin{equation}
 2(\theta + 1) + \delta(2\beta -3) > \gamma~,
\end{equation}
which results in the condition $\alpha(\alpha - 2\delta) < 0$ that is satisfied for $0 < \alpha < 2\delta$. Therefore, the conformon model with two symmetry-breaking parameters can facilitate the localization on the brane of both gravitational and gauge field zero modes.


\section{Localization of fermions}

In this section, we will apply the same method to localize the Dirac field on the brane. The initial attempt involves the action
\begin{equation}
  S = \int d^{5}x\sqrt{-g}\chi^{(\beta - 3/2)}\left[\bar{\Psi}\Gamma^{\mu}D_{\mu}\Psi -(D_{\mu}\bar{\Psi})\Gamma^{\mu}\Psi\right]\;,
\end{equation}
which leads to the equation of motion:
\begin{equation}
 2\chi^{(\beta - 3/2)}\gamma^{\mu}\partial_{\mu}\Psi(x,z) + 2 \chi^{(\beta - 3/2)}\gamma^{5}(\partial_{z} +2A')\Psi(x,z) + \Psi(x,z)\gamma^{5}\partial_{z}(\chi^{(\beta - 3/2)}) = 0\;.
\end{equation}
For the zero mode, $\gamma^{\mu}\partial_{\mu}\Psi(x,z) = 0$, we obtain the solution
\begin{eqnarray}
 g(z) \propto e^{- [2 + (\beta - 3/2)^2\alpha/2]A}~,
 \end{eqnarray}
where we used that $\chi = C e^{(\beta - 3/2)A}$ and $\Psi(x,z) = g(z)\tilde{\Psi}(x)$. Unfortunately, the above solution does not localize the zero mode of the Dirac field. 

Thus, the method of breaking the conformal symmetry can not provide the localization, so we will use this symmetry to build a mass term in the form:
\begin{eqnarray}\label{SDirac}
  S = \int d^{5}x\sqrt{-g}\left[i\bar{\Psi}\Gamma^{\mu}D_{\mu}\Psi +\lambda\chi^{-8/3}\bar{\Psi}\Psi\right]\;,
\end{eqnarray}
where $\lambda$ is a coupling constant. The above action leads to the equation of motion:
\begin{eqnarray}
i\gamma^{\mu}\partial_{\mu}\Psi + i\gamma^{5}\left(\partial_{5} + 2A'\right)\Psi + \lambda\chi^{-8/3}e^{A(z)}\Psi = 0 \;.
\end{eqnarray}
Performing the separation of the chiralities in the form:
\begin{equation} \label{sf}
 \Psi(x,z) = e^{-2A}\sum_{n}\left[\psi_{n}^{+}(x)\xi_{n}^{+}(z) +\psi_{n}^{-}(x)\xi_{n}^{-}(z)\right]\;,
\end{equation}
using that $i\gamma^{\mu}\partial_{\mu}\psi_{n}^{\pm}(x) = 0$ and $i\gamma^{5}\psi_{n}^{\pm}(x) = \mp\psi_{n}^{\pm}(x)$, the equation of motion leads to the solution:
\begin{equation}
\xi_{n}^{\pm}(z) \propto \mbox{exp}\left[\pm\lambda\int^{z}\chi(z')^{-8/3}e^{A(z')} dz'\right] ~.
\end{equation}
In this way, the condition of the localization can be written as:
\begin{equation}
   I_{\pm} = \int_{-\infty}^{\infty} dz\; \mbox{exp}\left[\pm 2k\lambda (k|z| +1)^{-2\left(-8\beta/3 +5\right)/\gamma +1}\right]~,
\end{equation}
where we had used the solution of the conformon (\ref{chi}) and the warp factor (\ref{b}). Replacing the values of $\beta$ and $\gamma$ as a function of the breaking symmetry parameter $\alpha$ from (\ref{beta,gamma}), we conclude that the zero modes of the Dirac field can be localized in the brane model presented in this paper for any $\alpha >0$. However, as in other models, only one chirality can be localized, depending on the sign of the Yukawa coupling constant $\lambda$.

The logarithm of the solution (\ref{chi}) can also be used to show the localization of Dirac fields within our model. Let us consider the action
\begin{equation}
    S = \int d^{5} x \sqrt{-g} \left[   \bar{\Psi} \Gamma ^{M}D_{M} \Psi + \lambda \bar{\Psi}\Gamma^{M} \partial_{M} \chi \gamma^{5} \Psi \right]~,
\end{equation}
which leads to the following equation of motion:
\begin{equation}
    \left\{\gamma^{\mu} \partial_{\mu} + \gamma^{5}\left( \partial_{z} + 2A'\right) + \lambda \partial_{z} \chi \right\} \Psi = 0~.
\end{equation}
Using similar to (\ref{sf}) solution, we find the equations for $\xi^{+}(z)$ and $\xi^{-}(z)$:
\begin{equation}
    \left(\partial_{z} + \lambda \partial \chi \right) \xi^{+}(z) = -m \xi^{-}(z)~, \qquad \left(\partial_{z} - \lambda \partial \chi \right) \xi^{-}(z)= m \xi^{+}(z)~.
\end{equation}
The solution for the zero mode of these equations is:
\begin{equation}
    \xi_{\pm}(z)= e^{\mp \int _{0}^{z} d \bar{z} \lambda \partial_{\bar{z}} \chi}~.
\end{equation}
If we take $ \chi \rightarrow \ln{\chi}$, we have:
\begin{equation}
    \xi_{\pm}(z)= e^{\mp (\beta - 3/2)\lambda  A} ~,
\end{equation}
where $ \ln{\chi} = (\beta - 3/2) A$. With that we get the following effective action:
\begin{equation}
    S = \int dz e^{-2 (\beta - 3/2)\lambda  A} \int d^{4} x \bar{\Psi}^{+} \gamma^{\mu} \partial_{\mu} \Psi^{+} + \int dz e^{2 (\beta - 3/2)\lambda  A } \int d^{4} x \bar{\Psi}^{-} \gamma^{\mu} \partial_{\mu} \Psi^{-}~.
\end{equation}
The left chirality is localized for $\lambda<0$. Choosing $\lambda=-1$ we get the range for the $\alpha$ parameter that localizes the field:
\begin{equation}
    -\frac{2}{3}\left(\sqrt{5}+5 \right) < \alpha < \frac{2}{3}\left(\sqrt{5}-5 \right)~.
\end{equation}
For the right chirality, we need $ \lambda > 0$. Choosing $ \lambda= 1$ we find:
\begin{equation}
    -\frac{4}{3}\left(\sqrt{11} + 4\right) < \alpha < \frac{4}{3}\left(\sqrt{11} -4 \right)~.
\end{equation}
These ranges for $\alpha$ indicate the conditions under which the left- and right-chiral components of the Dirac field can be localized on the brane within the presented model.


\section{Conclusion}

The study investigates a conformal 5D braneworld, focusing on the emergence of warping through the breaking of conformal symmetry. The model is characterized by including conformal breaking parameters, influencing the localization of different fields on the brane. It is established that the model, featuring a single conformal breaking parameter, successfully localizes gravity on the brane. However, it falls short in confining gauge bosons, necessitating the introduction of a second conformal breaking parameter to achieve localization of gauge field zero modes. When subjected to another symmetry breaking, the Dirac field fails to attain zero-mode localization. Two proposed methods are introduced to address the localization of Dirac field zero modes. The first method employs a Yukawa coupling, demonstrating the ability to achieve localization for any positive value of the conformal breaking parameter $\alpha$. The second method, involving the derivative of the logarithm of the conformon, yields negative values for $\alpha$. However, as observed in many brane models, only one chirality can be effectively localized, contingent on the sign of the coupling constant.


\section*{Acknowledgments}

We acknowledge the financial support provided by Funda{\c c}{\~a}o Cearense de Apoio ao Desenvolvimento Cient{\'i}fico e Tecnol{\'o}gico (FUNCAP), the Conselho Nacional de Desenvolvimento Cient{\'i}fico e Tecnol{\'o}gico (CNPq).


\end{document}